\documentstyle[multicol,epsfig,aps,prb]{revtex}
\ifpreprintsty \def\multb{ } \def\multe{ } \else \def\multb{ \begin{multicols}{2}} \def\multe{ \end{multicols}} \fi

\begin{document}

\title{Tunneling of Bloch electrons through vacuum barrier}
\author{I.I. Mazin}
\address{code 6391, Naval Research Laboratory, Washington, DC 20375, USA}
%\institute{code 6391, Naval Research Laboratory, Washington, DC 20375, USA}
%\pacs{73.40.Gk }{Tunneling}

\maketitle

\begin{abstract}
Tunneling of Bloch electrons through a vacuum barrier introduces 
new physical effects in comparison with the textbook case of free
(plane wave) electrons. For the latter, the exponential
decay rate in the vacuum
is minimal for electrons with the parallel component of momentum
${\bf k}_\parallel=0$, and the prefactor is defined by the electron
momentum component in the normal to the surface direction.
However, the decay rate of Bloch
electrons may be minimal at an arbitrary ${\bf k}_\parallel$
(``hot spots''), and the
prefactor is determined by the electron's group velocity,
rather than by its quasimomentum. 
\end{abstract}

\multb
There is a general feeling in the applied physics community nowadays that
the next decade will bespeak an advent of magnetoelectronics, exploiting
spin, rather than charge, degrees of freedom\cite{prinz}.
Many of such spintronic devices are based on the phenomenon of quantum
tunneling, and specifically on the difference between tunneling currents in
different spin channels. this opens up a
possibility to control the
electric properties {\it via} magnetic field. In view
of all that, an explosion of publication on spin-polarized tunneling which
started around 1995, is not surprising at all.

Interestingly, despite the fact that the tunnelling is one of the best studied
phenomena in quantum mechanics, there is still substantial diversity in
microscopic understanding of tunneling in real systems. Tunneling problems
are very easily solved in one dimension and for free electrons;
but it is not so obvious  how this is related to real systems, and how to
incorporate the effects of realistic electronic structure.
This is the reason why most theoretical papers use the free-electron
model, that is, the electronic wave functions that are plane waves and a
spherical Fermi surface. The deviations of the wave function from the
single plane wave form, and of the Fermi surface from a sphere, are very often
crucial for understanding the physics of tunneling. In particular, 
a totally counterintuitive result has been observed in some recent
calculations\cite{JI}, when the electrons with nonzero quasimomentum
parallel to the interface had a larger probablility to tunnel through
a vacuum barrier compared to those with zero parallel quasimomentum.
It was also pointed out\cite{WB} that electrons in different Bloch 
states with the same energy  and quasimomentum may, in principle,
have different decay rates in vacuum: another counterintuitive
result. In this regards, it is important to establish a formal
theory for tunneling of Bloch electrons through a vacuum barrier,
elucidating the qualitatively new aspects of this process as opposed to
the free electron tunneling, particularly because only the latter
is discussed in the classical textbooks. This is the goal
of the current  paper.

The simplest case of a tunneling contact is the so-calles Sharvin contact%
\cite{sharvin}, which is essentially an orifice between two metals
(or a metal and the vacuum), whose
size is smaller than the mean free path of electrons in the bulk. All
electrons with the a positive projection onto the current direction (which
we will denote as 
$x$) pass through the contact. Conductance of a Sharvin contact
between two identical metals is 
\begin{equation}
G=\frac{e^{2}}{\hbar }\frac{1}{2}\left\langle N|v_{x}|\right\rangle A,
\label{ball}
\end{equation}%
where $A$ is the contact area, $N$ is the volume density of electronic
states at the Fermi level, $v$ is the Fermi velocity, and brackets denote
Fermi surface averaging: 
\multe
\begin{equation}
\frac{1}{2}\left\langle N|v_{x}|\right\rangle =\frac{1}{\Omega }\sum_{{\bf k}%
i\sigma }\delta (\epsilon _{{\bf k}i\sigma }-E_{F})v_{{\bf k}i\sigma ,x}=%
\frac{1}{(2\pi )^{3}}\sum_{i\sigma }\int \frac{dS_{F}}{|v_{{\bf k}i\sigma }|}%
v_{{\bf k}i\sigma ,x}.  \label{sharvin}
\end{equation}%
\multb
Integration and summations are over the states with $v_{{\bf k}i\sigma ,x}>0,
$ and $\Omega $ is the unit cell volume\cite{schep,im99}. ${\bf k},$ $i,$ and 
$\sigma $ denote the quasimomentum, the band index, and the spin of an
electron, respectively. This formula can be derived by considering the
voltage-induced shift of the Fermi surface\cite{im99}, but there is a more
instructive derivation starting from the Landauer-Buttiker formula for the
conductance of a single ballistic electron, $G_{0}=e^{2}/h.$ In this
formalism, the total conductance is equal to $G_{0}$ times the number of
conductivity channels, $N_{cc}$, which is defined as the number of electrons
that can pass through the contact. Assuming that the translational symmetry
in the interface plane is not violated, we observe that the quasimomentum in
this plane, {\bf k}$_{\parallel },$ is conserved, and $N_{cc}$
is the number of quantum-mechanically 
allowed ${\bf k}_{\parallel }$'s.
Thus $N_{cc}$ is given by the total area of the contact times the density of the
two-dimensional quasimomenta.
 The latter is simply $S_{x}/(2\pi )^{2},$ where $S_{x}$ is the
area of the projection of the bulk Fermi surface onto the contact plane.
Thus 
\begin{equation}
G= \frac{e^{2}}{h}\frac{S_{x}A}{(2\pi )^{2}}\equiv \frac{e^{2}}{%
\hbar }\frac{1}{2}\left\langle N|v_{x}|\right\rangle A .
\label{proj}\end{equation}
This is an important result. To the best of our knowledge, Walter Harrison
was the first to spell it out in 1961\cite{walter}, and there is no lack of
more recent papers manifesting proper understanding of this issue (e.g., Ref.%
\cite{yip}). However, till now many otherwise correct and useful papers
erroneously identify the number of conductivity channels and the density of
states at the Fermi level, that is, 
\multe
\begin{eqnarray}
\frame{$N_{cc}\propto N(E_{F})$} &=&\frac{1}{\Omega }\sum_{{\bf k}i\sigma
}\delta (\epsilon _{{\bf k}i\sigma }-E_{F})=\frac{1}{(2\pi )^{3}}%
\sum_{i\sigma }\int \frac{dS_{F}}{|v_{{\bf k}i\sigma }|}. \\
incorrect! &&  \nonumber
\end{eqnarray}

\multb
Eq. \ref{proj} is the basis for all more sophisticated expressions
describing various aspects of quantum tunneling. None of
them may explicitely depend on the bulk density of states. It may, however,
be that that instead of the straight $\left\langle N|v_{x}|\right\rangle $
averaging one has to compute a weighted average, with the weights coming
from tunneling matrix elements, or other additional physics.

Eq. \ref{proj} takes care of one important difference between the
free electrons and the Bloch electrons: deviation of the Fermi
surface from a sphere, for $S_x\neq \pi k_F^2$. Another important
difference that is often neglected
 is that between the group velocity $\hbar^{-1}d
 \epsilon_{\bf k}/d{\bf k}$
and the phase velocity ${\hbar\bf k}/m_0$. 
One can get some qualitative understanding
 of the role that this fact plays in tunneling
by considering 
a simplified model, where electrons in metal
are approximated by free electrons with an effective mass different from the
free electron mass. In this approximation, the effective mass is responsible
for the difference 
between  ($\hbar {\bf k}/m_{0})$ and 
($d\epsilon _{{\bf k}}/\hbar d{\bf k)}$, and the
transparency of a symmetric rectangular barrier is defined by the standard
formula (see, e.g., Ref.\cite{Landau}): 
\multe
\begin{equation}
D({\bf k})=\frac{4m_{0}^{2}\hbar ^{2}K^{2}v_{L}v_{R}}{\hbar
^{2}m_{0}^{2}K^{2}(v_{L}+v_{R})^{2}+(\hbar
^{2}K^{2}+m_{0}^{2}v_{L}^{2})(\hbar ^{2}K^{2}+m_{0}^{2}v_{R}^{2})\sinh
(dK)^{2}},  \label{D}
\end{equation}%
\multb
where $v_{L(R)}$ stands for ({\bf k}-dependent) Fermi velocity in the left
and in the right leads,
and the imaginary
quasimomentum $\hbar K$ is calculated from the energy conservation
condition, \[ U+\hbar ^{2}[{\bf k_\parallel}^{2}-K^{2}]/2m_{0}=E, \]%
where $m_{0}$ is the free electron mass, $U$ is the barrier height,
and $d$ is its thickness.

The physical reason that one has to
use group velocities, and not wave vectors, is very profound
and extends well beyound the limited scope of the effective mass 
model: these factors appear in the Eq.\ref{D} as a result
of matching the gradients of the wave functions at the interface,
and the gradient is, in fact, the velocity operator for the Bloch waves.
Another way to express the same idea is to recall the physical meaning of
 the usual quantum-mechanical
 requirement
that the wave functions be smooth: it is needed to ensure the flux continuity
and, therefore, particle conservation. On the other hand, the expression
for $K$ includes the {\it momentum}, ${\bf k_\parallel ,
}$, and the {\it free} electron
mass, $m_0$, because it comes from the solution of the Schr\"odinger
equation inside the barrier (in vacuum)\cite{Noz}.
The conductance of a contact described by Eq. \ref{D} is given by
the appropriately modified Eq.\ref{ball}: 
\begin{equation}
G=\frac{e^{2}}{\hbar }\frac{A}{\Omega }\sum_{{\bf k}}\delta (\epsilon _{{\bf %
k}}-E_{F})v_{{\bf k}x}D({\bf k}),  \label{D1}
\end{equation}

It is instructive to consider the last formula in some limiting cases.
First, let us consider a {\it specular} barrier. It is defined by the
limit $U\rightarrow \infty ,$ $d\rightarrow 0,$ $Ud=V.$ Then $K\rightarrow 
\sqrt{2m_{0}U/\hbar ^{2}},$ and 
\begin{equation}
D({\bf k})=\frac{4\hbar ^{2}v_{L}v_{R}}{\hbar ^{2}(v_{L}+v_{R})^{2}+4V^{2}},
\label{D2}
\end{equation}%
Note that in the literature the ratio $V/\hbar v_{x}=Z$ is commonly used to
characterize the barrier strength. In principle, this quantity is different
for different electrons, as $v_{x}$ depends on ${\bf k.}$ In the limit of
low transparency, $Z\gg 1,$ $D({\bf k})=\hbar ^{2}v_{L}v_{R}/V^{2}.$
Substituting this into Eq.\ref{D1}, we find that the total current is
proportional to%
\begin{equation}
\sum_{{\bf k}}\delta (\epsilon _{{\bf k}}-E_{F})v_{{\bf k}x}v_{L}v_{R},
\label{D4}
\end{equation}%
where summation is, of course, over those {\bf k} that are allowed in both
left and right lead. Roughly speaking, the total conductance is defined by
the smaller of the two $\left\langle Nv_{x}^{2}\right\rangle $'s, that is,
by $\min \left (
\langle Nv_{x}^{2}\rangle _{L},\langle
Nv_{x}^{2}\rangle _{R}\right )$. In the high transparency limit $D$ is still
smaller than 1, $D=4v_{L}v_{R}/(v_{L}+v_{R})^{2},$ (so-called Fermi velocity
mismatch), but in most cases this is not a large effect: factor of two mismatch
reduces $D$ by only 10\%. 

In the case of a thick barrier, defined as $dK\gg 1,$ Eq. \ref{D} can
be expanded in $\hbar ^{2}k_{\parallel }^{2}/4m_{0}(U-E_{F}),$ and the
transparency is 
\multe
\begin{equation}
D({\bf k})=\frac{2m_{0}^{2}(U-E_{F})v_{L}v_{R}}{%
(U-E_{F}+m_{0}v_{L}^{2}/2)(U-E_{F}+m_{0}v_{R}^{2}/2)}\exp (-2d^{2}W)\exp [-%
\frac{k_{\parallel }^{2}}{W}],  \label{D3}
\end{equation}%
where $\sqrt{2m_{0}(U-E_{0})}/\hbar d=W\ll k^{2}$ (thick barrier limit). $W$
does not depend on ${\bf k.}$ The tunneling current is proportional to 
\begin{equation}
J\propto \sum_{{\bf k_\parallel}}\frac{2m_{0}^{2}(U-E_{F})v_{L}v_{R}
}{(U-E_{F}+m_{0}v_{L}^{2}/2)(U-E_{F}+m_{0}v_{R}^{2}/2)}\exp [-%
\frac{k^{2}}{W}]
\end{equation}

Let ${\bf k}_{n}$ be the set of points on the Fermi surface where ${\bf k}_
\parallel =0$
(note that for Bloch electrons beyond the effective mass approximation
 tunneling from some of these  points may be
 suppressed by symmetry, as discussed
later in the paper). Except in the exponent, we can put ${\bf k}\parallel
$ to zero,
\begin{eqnarray}
J &\propto &\frac{1}{(2\pi )^{3}}\sum_{n}\left\{ \int d^{2}k\exp [-\frac{%
k^{2}}{W}]\right\} \frac{%
2m_{0}^{2}(U-E_{F})v_{L}v_{R}}{%
(U-E_{F}+m_{0}v_{L}^{2}/2)(U-E_{F}+m_{0}v_{R}^{2}/2)}  \label{thick} \\
&\propto &\sum_{n}\frac{v_{L}}{m_{0}\hbar ^{2}v_{L}^{2}/2+U-E_{F}}\frac{v_{R}%
}{m_{0}\hbar ^{2}v_{R}^{2}/2+U-E_{F}}.\nonumber
\end{eqnarray}%
\multb
All omitted in this expression factors are  ${\bf k}\parallel$-independent.
One should not be confused by the fact that, unlike Eq. \ref{D4}, the
numerator here does not have the third velocity. We have reduced our problem
to an effective 1D problem, in which case the role of the density of states
is played by the inverse velocity. Correspondingly, the product $Nv$ cancels
out.

Eqs. \ref{D4},\ref{D3} emphasize the role of kinematics in tunneling. For
instance, the long-standing problem of the reversed (compared to the density
of states) spin polarization of the 3d ferromagnets is entirely explained in
terms of kinematics. Direct calculations show that $s$-like electrons in Fe,
Co and Ni have much larger Fermi velocity than $d$-like electrons. Taking
this fact into account brings the calculated spin polarization to a very
good agreement with experiment, without making any additional assumptions
about the character of the surface states\cite{im99,NiFe}. This is by no mean
surprising: the bulk transport is controlled by the same factor $%
\left\langle Nv_{x}^{2}\right\rangle ,$ and the Ohmic current in these
metals is carried predominantly by  $s$-like electrons. It is only natural
that in another transport phenomenon, tunneling, these electrons also play
the leading role. We would like to emphasize that the effect considered above
(as opposed to another effect discussed later in the paper) is not related
to the $s$ or $d$ symmetry of the wave functions, but to the group velocities
in the respective bands. In other cases the ``light'' and the ``heavy'' bands
may not be directly related to the angular symmetry of the wave functions. For
example, in SrRuO$_3$ both spin-up and spin-down Fermi surfaces are made up
by Ru $t_{2g}$ $d$-electrons, but the average group velocity in the spin-majority
 channel is twice smaller than that in the spin-minority one\cite{david}. As
a result, although the spin polarization of the density of states is positive, $N_
\uparrow >N_\downarrow$, while the transport spin polarization is negative, $
\langle Nv\rangle_ \uparrow <\langle Nv\rangle_\downarrow$.\cite{Wor}

Now we have some understanding of the two remarkable differences between
the free electrons and the Bloch electrons: the effect of the
Fermi surface geometry and the difference between the group and the phase
velocities. There is, however, yet another, extremely important,
dissimilarity between the two systems, recently pointed out by 
W. Butler\cite{WB}: the difference between the momentum and
the quasimomentum. In order to discuss this difference, and its physical 
consequences, let us
consider reflection of an individual Bloch wave from a metal surface.
Let $x$ be the
direction normal to the surface, and {\bf r } the coordinate in the surface
plane. At $x<0$
 we have a metal, and vacuum at $x>0$. The vacuum potential is again
$U,$ and the
Fermi energy is $E.$ Since we have perfect in-plane
periodicity, the wave
function at any $x$ can be classified by $\bf k_\parallel$,
 and is given by%
\begin{equation}
\psi ({\bf k_\parallel},x,{\bf r_\parallel})=\sum_{{\bf G}}\exp [i({\bf k_\parallel
+G}){\bf r_\parallel}]F_{{\bf G}}(%
{\bf k_\parallel,}x).  \label{1}
\end{equation}%
The quasimomentum in the surface plane, $\hbar {\bf k_\parallel}$, is conserved, as
well as the energy. In vacuum, the solution of the Schr\"odinger equation is 
\begin{equation}
\psi ^{T}({\bf k_\parallel},x,{\bf r_\parallel})=
\sum_{{\bf G}}\alpha _{{\bf G}}\exp [i({\bf k_\parallel+G}%
){\bf r_\parallel}]\exp (-K_{{\bf G}}x),  \label{psig}
\end{equation}%
where {\bf G} is the 2D reciprocal lattice vector,
and $K_G$ is now defined taking into account the kinetic energy
associated with the given reciprocal lattice vector,
$ U+\hbar ^{2}[{\bf k_\parallel}^{2}-K_{\bf G}^{2}]/2m_{0}=E.$ 
 An incoming Bloch wave with a given 
${\bf k}$ penetrates into the barrier as a linear combination (\ref{psig})
with the coefficients $\alpha _{{\bf G}}$ defined by matching conditions,
set by the requirement of continuity of the wave function and its
derivative: 
\begin{eqnarray*}
\sum_{{\bf G}}F_{{\bf G}}(0)\exp [i({\bf k_\parallel
+G}){\bf r_\parallel}] &=&\sum_{{\bf G}%
}\alpha _{{\bf G}}\exp [i({\bf k_\parallel +G}){\bf r_\parallel}] \\
\sum_{{\bf G}}F_{{\bf G}}^{\prime }(0)\exp [i({\bf k_\parallel
+G}){\bf r_\parallel}] &=&-\sum_{%
{\bf G}}\alpha _{{\bf G}}K_{{\bf G}}\exp [i({\bf k_\parallel
+G}){\bf r_\parallel}].
\end{eqnarray*}%
since this has to hold for any ${\bf r_\parallel},\ \alpha _{{\bf G}}=F_{{\bf G}}(0)$
, and $F_{{\bf G}}^{\prime }(0)=-\alpha _{{\bf G}}K_{{\bf G}}$ for each $%
{\bf G.}$ Thus 
\begin{equation}
F_{{\bf G}}(0)K_{{\bf G}}+F_{{\bf G}}^{\prime }(0)=0.  \label{all}
\end{equation}%
If $F_{{\bf G}}(x)$ were a linear combination of the bulk Bloch waves with
the energy $E$ and the quasimomentum in the plane $\hbar {\bf k_\parallel ,}$ 
\begin{equation}
F_{{\bf G}}(x)=u_{k_{x}}(x)\exp (ik_{x}x)+au_{-k_{x}}(x)\exp (-ik_{x}x),
\label{bulk}
\end{equation}%
we would have only one free parameter, $a,$ to satisfy Eq.\ref{all} for all 
{\bf G}'s, which is obviously impossible. The answer is that $F_{{\bf G}}(x)$
has the form (\ref{bulk}) only far away from the surface, while near the
surface it is distorted as required by Eq.\ref{all}. This emphasizes once again
the role of of surface states in tunneling. In fact, one of the ways to
realize the necessity of forming the surface states is that the bulk Bloch
functions, in general, cannot be augmented continuously and smoothly into
vacuum.

In the case of a thick barrier, the actual tunneling current will be defined
by that component of the wave function (\ref{psig}) which has the smallest $%
K,$ that is, by the one with ${\bf G=0.}$ The amplitude of this evanescent
wave is set by $\alpha _{{\bf 0}}.$ 
As pointed out by Butler\cite{WB},
${\bf k_\parallel =0}$ is
a high symmetry direction ($\Gamma $X), and the electronic states possess
certain symmetry in the $yz$ plane. In particular, $\alpha _{{\bf 0}}$ for
some states may vanish by symmetry, in which case the decay rate $K$ will be
defined by the smallest $G$ allowed by symmetry. Since we consider now a
thick barrier, this essentially means that tunnelling from such a band will
be defined not by the ${\bf k_\parallel =0}$ state, but, rather
counterintuitively, by general (not high symmetry)
points in the 2D Brillouin zone (as confirmed by actual
calculations\cite{JI}).
 Indeed, consider a band where
by symmetry $F_{0}(0,x)=0$ at ${\bf k_\parallel =0.}$ At ${\bf k_\parallel
\neq 0}$ thus $F_{0}(%
{\bf k_\parallel},x)=$ $F_{0}^{\prime \prime }({\bf 0},x)k_\parallel
^{2},$ while $K=\sqrt{%
2m_{0}(U-E)/\hbar ^{2}+k^{2}}\approx \sqrt{2m_{0}(U-E)}/\hbar +\hbar k_\parallel
^{2}/2%
\sqrt{2m_{0}(U-E)}=K_{0}+k_\parallel^{2}/2K_{0}.$ The optimal distance from the zone
center that gives maximal contribution to the tunneling current can be
estimated by maximizing with respect to $k_\parallel$ of 
\[
F_{0}^{\prime \prime }({\bf 0},x)k_\parallel^{2}\exp (-K_{0}d-k_\parallel
^{2}d/2K_{0}),
\]%
where $d$ is the barrier thickness, which gives $k_\parallel\sim \sqrt{2K_{0}/d}%
.$ For Fe, for instance, $K_{0}\approx 0.6$ a.u., about the same as the $%
\Gamma $X distance. Thus for a barrier, say, of 5 lattice parameters, $k
_\parallel%
\sim 0.2$ a.u., a sizeable distance from the center of the Brillouin
zone. Yet another counterintuitive result is that the low transparency
limit is not unique:
 in the thick barrier limit tunneling is
predominantly from the states infinitely close to the zone center.
However, in the high barrier limit, which is another way to implement
a low transparency asymptotics, tunneling occurs far away from the zone center, 
possibly at the zone boundary. This is the effect observed in Refs.\cite{JI}.

To conclude, we discussed here three new effects which appear in tunneling
of the Bloch electrons through a vacuum barrier, as compared with
the textbook case of free electron (plane wave) tunneling.
These effects are due to (i) complexity of the Fermi surface geometry
(``fermiology''),
(ii) difference between the group and the phase velocities of a
Bloch electron, and (iii) nonconservation of the parallel component
of electron momentum (and conservation of its quasimomentum). Each
effect influences the tunneling current in its own way, and as a result
even for the most simple case of a vacuum barrier, the tunneling
of the Bloch electrons appears to be qualitatively different 
from the free electron tunneling.
\acknowledgments
Discussions with J. Kudrnovsky, I. Mertig, and especially
with B. Nadgorny and W. Butler, are gratefully 
acknowledged.

 \multe

\end{document}